\documentclass[12pt]{article}
\usepackage{fullpage,amsmath}
\usepackage[pdftex,dvipsnames,usenames]{color}         
\usepackage{xcolor}
\usepackage{natbib,bm}
\usepackage{float}
\usepackage{lscape}
\usepackage{bbm}
\usepackage{graphicx,psfrag,epsf}
\usepackage{enumerate}
\usepackage{enumitem}
\usepackage{natbib}
\usepackage{bbm} 
\usepackage{amsmath,amsthm,amssymb,bm}
\usepackage{multirow}

\usepackage{abstract}

\usepackage{amsthm} 

\newcommand{\comment}[1]{}

\def\boldfacefake#1{\kern-4pt
   \hbox{ \mathsurround=0pt
   \hbox to 0.4pt{$#1$\hss}\hbox to 0.4pt{$#1$\hss}\hbox {$#1$}}}

\newcommand{\ba}{\begin{eqnarray*}}
\newcommand{\ea}{\end{eqnarray*}}

\newtheorem{theorem0}{Theorem}
\newtheorem{lemma0}{Lemma}
\newtheorem{remark0}{Remark}
\newtheorem{fact0}{Fact}
\newtheorem{example0}{Example}
\newtheorem{definition0}{Definition}
\newtheorem{corollary0}{Corollary}
\newtheorem{proposition0}{Proposition}
\newtheorem{algorithmY}{Algorithm}
\newtheorem{condition0}{Condition}
\newtheorem{assumption0}{Assumption}
\newtheorem{simulation0}{Simulation}

\newenvironment{lemma}{\begin{lemma0} \mbox{}}{\end{lemma0}}

\newcommand{\reals}{\mbox{\rm I\kern-.20em R}}
\newcommand{\sreals}{\mbox{\small \rm I\kern-.20em R}}

\ifx\pdfoutput\undefined 
\usepackage[dvips]{graphicx}
\else
\usepackage{epstopdf}

\newcommand{\bqg}{\begin{quote} \color{Green}\em}
	\newcommand{\toale}{\end{quote} \color{black}\rm}
\newcommand{\eqg}{\end{quote} \color{black}\rm}

\newcommand{\bd}{\begin{description}}
\newcommand{\ed}{\end{description}}
\newcommand{\bi}{\begin{itemize}}
\newcommand{\ei}{\end{itemize}}
\newcommand{\be}{\begin{enumerate}}
\newcommand{\ee}{\end{enumerate}}

\newcommand{\pkg}[1]{{\fontseries{b}\selectfont #1}}

\newcommand\blfootnote[1]{%
  \begingroup
  \renewcommand\thefootnote{}\footnote{#1}%
  \addtocounter{footnote}{-1}%
  \endgroup
}

\setcitestyle{square}

\begin{document}

		\begin{center}
		{\large \bf A Bias Correction Method in Meta-analysis of Randomized Clinical Trials with no Adjustments for Zero-inflated Outcomes
		}
		
		\vspace{.5 cm}
		
		Zhengyang Zhou, Ph.D. \blfootnote{Correspondence should be sent to: Zhengyang Zhou, Ph.D. (Email: zhengyang.zhou@unthsc.edu) and Eun-Young Mun, Ph.D. (Email: eun-young.mun@unthsc.edu)}\\
		University of North Texas Health Science Center, Fort Worth, TX \\
		
		Minge Xie, Ph.D. \\
		Rutgers University,
		Piscataway, NJ \\

		David Huh, Ph.D.  \\
		University of Washington,	Seattle, WA\\

		Eun-Young Mun, Ph.D.  \\
		University of North Texas Health Science Center, Fort Worth, TX \\
		
	\end{center}

	\begin{abstract}
		\begin{center}
		{\large \bf Summary
		}
		\end{center}
	Many clinical endpoint measures, such as the number of standard drinks consumed per week or the number of days that patients stayed in the hospital, are count data with excessive zeros. However, the zero-inflated nature of such outcomes \textcolor{black}{is sometimes ignored in analyses of clinical trials. This leads to biased estimates of study-level intervention effect} and, consequently, a biased estimate of the overall intervention effect in a meta-analysis. The current study proposes a novel statistical approach, the Zero-inflation Bias Correction (ZIBC) method, that can account for the bias introduced when using the Poisson regression model, despite a high rate of \textcolor{black}{inflated} zeros in the outcome distribution of a randomized clinical trial. This correction method \textcolor{black}{only requires} summary information from individual studies to correct intervention effect estimates as if they were appropriately estimated using the zero-inflated Poisson regression model, \textcolor{black}{thus it is attractive for meta-analysis when individual participant-level data are not available in some studies.}  Simulation studies and real data analyses showed that the ZIBC method performed well in correcting zero-inflation bias in most situations. 
	\end{abstract}
	

\section{Introduction}
\label{sec:intro}

Meta-analysis is an established statistical approach for combining data from multiple studies to provide large-scale evidence across many disciplines, including medical, educational, and policy research \citep{schmid2020handbook}.  The majority of published meta-analyses have relied on aggregate data (AD), which are study-level summary statistics available from published or unpublished reports  \citep{sutton2008recent, lyman2005strengths, chen2020relative}.
However, AD meta-analysis is susceptible to estimation bias, because the biased result from a study with model misspecification
 (e.g., a biased effect size) will be carried over in meta-analysis if the study is included.  For AD meta-analysis, it is challenging to correct biased estimation from original studies without refitting raw individual participant data (IPD) using a more suited statistical model \citep{liu2018avenues}. In this paper, we aim to correct this estimation bias, i.e., the bias from the conventional count model on zero-inflated count outcome, when only AD are available for meta-analysis. 

Count outcomes are prevalent in clinical research, including number of seizures for each patient in epilepsy trials (e.g., \cite{garcia2004trial}), number of relapses in multiple sclerosis trials (e.g., \cite{silcocks2010p3mc}), and number of standard alcohol drinks in alcohol intervention trials (e.g., \cite{huh2019tutorial}). Some studies, by nature, have high proportions of zero outcome values. \textcolor{black}{For example, drinking outcomes from alcohol intervention studies reflect both individuals who abstain from alcohol following an intervention and those who happened not to drink, resulting in a large proportion of zero drinks, above and beyond the frequency that would be predicted by conventional count models, such as the Poisson.} Therefore, estimation results \textcolor{black}{might} be biased if the Poisson regression model was used in studies with \textcolor{black}{potentially} zero-inflated outcomes \textcolor{black}{(\cite{horton2007cautionary}; some examples may be \cite{ milgrom2009xylitol, frazao2011effectiveness, kelly2017pilot}). If the results from Poisson regression are included in a meta-analysis, the biased estimation results might} further bias the pooled result in a meta-analysis. We, henceforth, refer to this bias as
\textit{zero-inflation bias} throughout the study.  \textcolor{black}{The application of appropriate statistical approaches to accommodate zero-inflated outcome data has been on the rise in recent years, with the availability of relevant software packages (e.g., zero-inflated and hurdle models in the \pkg{pscl} R package \citep{zeileis2008regression}). 
However, there still exists a non-ignorable number of publications that did not ideally account for zero-inflation in analysis. For example, in a recent meta-analysis study of 17 brief alcohol interventions, the proportions of participants reporting zero number of drinks were considerably high in nine studies, eight of which did not account for zero-inflation in outcome reporting  \citep{huh2015brief}. The studies that did not properly account for zero-inflation may still be pooled in meta-analysis studies for years to come. Thus, a methodological approach capable of correcting biased estimates from past studies would help facilitate AD meta-analyses of zero-inflated count outcomes into the future.}


A zero-inflated Poisson (ZIP) model is more appropriate for count data with many zeros, since it assumes that the outcome follows a mixture of a point mass at zero and a Poisson distribution \citep{lambert1992zero}. From a clinical perspective, the two components of the ZIP model correspond to two distinct subpopulations: (a) participants who predictably do not engage in the behavior, and (b) participants who may or may not engage in the behavior at a particular assessment. In some clinical situations, clinicians may focus on the latter as they are the primary target of their intervention (\textcolor{black}{e.g., whether an alcohol intervention helps reduce drinking for those who regularly engage in drinking; } See Section 2.1 for two examples). In this paper, \textcolor{black}{we are interested in the incidence density ratio for intervention on the mean of the Poisson portion in the ZIP model, which is important for understanding the intervention effect among the subpopulation that may potentially engage in the behavior.}
Note that in other trial evaluation situations, \textcolor{black}{modeling the overall mean of the outcome, which accommodates structural zeros, may be more desirable   \citep{long2014marginalized}.}
 
In this article, we focus on  mitigating the impact of zero-inflation bias in meta-analysis and propose a novel statistical method, called a Zero-inflation Bias Correction (ZIBC) method. This method corrects the biased intervention effect size estimation that can result from the conventional Poisson regression model, the ``go-to" method when modeling count outcomes. We aim to correct zero-inflation bias and produce a bias-corrected effect size estimate equivalent to the estimate from the ZIP regression model. This bias correction is achieved by comparing the estimating equations under the ZIP and Poisson models and using summary statistics of intervention and control subgroups. We will refer to the Poisson and ZIP regression models as the \textit{conventional} and \textit{true} methods, respectively, in the current paper. 


The paper proceeds as follows. In Section 2, we describe the formulation of the standard Poisson and ZIP regression models for a single study. We then introduce the ZIBC method for correcting zero-inflation bias as well as how to apply it in an AD meta-analysis. In Section 3, we conduct simulation studies to evaluate the performance of the ZIBC method in bias correction.
 In Section 4, \textcolor{black}{we consider two real data examples. In the first example, we examine the intervention effects on alcohol use utilizing data drawn from an IPD meta-analysis study. The example is used to demonstrate the performance of the ZIBC method, pretending we have only AD, which were derived from IPD.  
 In the second example, we illustrate the application of the method in a clinical trial for preventing dental caries, utilizing AD from the published report. In this example, we have only AD and it is not possible to perform a meta-analysis using IPD.  
In Section 5, we discuss the overall findings and conclusions.} 

\section{Method: From single study to meta-analysis}
\label{sec:meth}
In this section, we describe the ZIBC method and how it corrects zero-inflation bias in an AD meta-analysis. 
We first focus on the case of a single randomized clinical trial, where we set up notations for the true and conventional methods (Section 2.1). We then describe zero-inflation bias (Section 2.2), and provide the ZIBC method that can correct it (Section 2.3). 
Next, for each clinical trial that originally used the conventional method for zero-inflated outcomes, we implement the ZIBC method to obtain the bias-corrected intervention effect estimate and conduct a standard meta-analysis for the overall bias-corrected intervention effect (Section~2.4).
\subsection{Model setup: Single randomized clinical trial}
\subsubsection{True method: ZIP regression model}
For a randomized clinical trial with two arms, we assume a count outcome with an excessive rate of zeros that follows a ZIP regression model. Suppose the study sample size is $n$, and for $i$-th subject, $i=1,2,...,n$, and we assume that the outcome $y_i$ is distributed
\begin{equation}
\label{eq:ZIP}
  y_i\sim \begin{cases}
               0 &\text{with probability } \pi_i\\
               \text{Poisson($\mu_i$)} &\text{with probability }1-\pi_i,
            \end{cases}
\end{equation}
where $\pi_i$ is the structural zero rate and $\mu_i$ is the mean parameter of the Poisson portion for subject $i$. \textcolor{black}{The mean of $y_i$ is $ \mathbb{E}[y_i]=(1-\pi_i)\mu_i$.}

In the context of \textcolor{black}{intervention or prevention studies}, the structural zeros correspond to participants that do not engage in the outcome (e.g., alcohol abstainers who do not drink across situation and time), whereas the Poisson portion corresponds to those who may or may not engage in the behavior at a given time or situation (e.g., participants who may or may not drink during the past month at 1-month follow-up). 
The present paper focuses on the Poisson portion characterizing the intervention effect on the latter, which is of interest  in many \textcolor{black}{harm-reduction alcohol intervention studies.} For example, in alcohol prevention and intervention trials among college students, researchers may be most interested in students who may drink if given an opportunity  (e.g., Section 4.1). Another example is clinical trials to prevent dental caries among children, where the outcome of interest is number of caries developed during a certain period (e.g., Section 4.2). Among the trials, some children may be unlikely to develop dental caries (e.g., due to good oral hygiene habits or protective genetic factors), while others have higher chances of developing them. Therefore, targeting the latter group of children, which can be characterized through the Poisson portion, may produce higher cost-effectiveness and utility for dental caries prevention strategies.


The Poisson portion can be modeled as follows. Suppose $p-1$ covariates are included in the model and one of the covariates is the intervention assignment indicator $\mathbbm{1}_{\{A_i=T\}}$, where $A_i$ denotes a participant's assignment to either the intervention ($T$) or control ($C$) arm, and ${\bf x}_{i,p-2}=(x_{i2},x_{i3},...,x_{i,p-1})^t$ denotes the remaining $p-2$ covariates. The Poisson mean parameter is estimated by the covariates in
\begin{equation}
\label{eq:logmu}
\log(\mu_i)={\bf x}_i^t{\bm \beta}=\beta_0+\beta_1\mathbbm{1}_{\{A_i=T\}}+{\bf x}_{i,p-2}^t{\bm \eta},
\end{equation}
where ${\bf x}_i=(1, \mathbbm{1}_{\{A_i=T\}}, {\bf x}_{i,p-2}^t)^t$ and ${\bm \beta}=(\beta_0,\beta_1,\bm \eta^t)^t=(\beta_0,\beta_1,\beta_2,...,\beta_{p-1})^t$ are the regression coefficients. Note that $\beta_1$ measures the intervention effect, \textcolor{black}{the
log incidence density ratio difference between the intervention and control groups, which is the parameter we aim to recover.} We denote $\bm \beta^0=(\beta_0^0,\beta_1^0,\bm \eta^{0,t})^t$ as the true regression parameters. 

From Equations (\ref{eq:ZIP}) and (\ref{eq:logmu}), the estimating equations under the true method is given by
\begin{equation}
\nonumber
S_\text{ZIP}(\bm \beta)  \triangleq
 \frac1n\sum_{y_i=0}\left[\frac{\pi_i}{\pi_i+(1-\pi_i)\exp\{-\exp({\bf x}_i^t{\bm \beta})\}}\right]\exp({\bf x}_i^t{\bm \beta}){\bf x}_i +  \frac 1n \sum_{i=1}^n\{-\exp({\bf x}_i^t{\bm \beta})+y_i \}{\bf x}_i.
\end{equation}
By solving $S_\text{ZIP}(\bm \beta)=0$, we can obtain the maximum likelihood estimates (MLE), $\hat{\bm \beta}_{\text{MLE}}$. As $\bm \beta^0$ are the true parameters for the ZIP model (\ref{eq:ZIP}),
 we also have $
\mathbb{E}[S_\text{ZIP}(\bm \beta^0)]=0$ and $\hat{\bm \beta}_\text{MLE} \to \bm \beta^0$ as $n \to \infty$ by standard likelihood inference. 
Note that $\pi_i$ can be modeled separately in a logistic model. However, we do not attempt to model $\pi_i$ because it is not the interest of the current study. 

\subsubsection{Conventional method: Standard Poisson regression model} 
For the same trial design described in the previous subsection, \textcolor{black}{some} researchers (cf., \cite{frazao2011effectiveness, murphy2000comparative}, etc.) have used the conventional (CV) Poisson model to analyze \textcolor{black}{potentially} zero-inflated count outcome $y_i$ with $
f_\text{CV}(y_i\,|\,\mu_i) =\frac{e^{-\mu_i}\mu_i^{y_i}}{y_i!}$, where $ \log(\mu_i) ={\bf x}_i^t{\bm \beta}$.

Under the conventional method, 
we derive the following estimating equations
\begin{equation}
\label{eq:Scv}
S_\text{CV}(\bm \beta)\triangleq 
\frac 1n \sum_{i=1}^n\{-\exp({\bf x}_i^t{\bm \beta})+y_i \}{\bf x}_i
\end{equation}
and denote $\hat{\bm \beta}_\text{CV}$
as the solution of $S_\text{CV}(\bm \beta)=0$. Then $\hat{\bm \beta}_\text{CV}$ are the parameter estimates in the conventional method, which are  usually reported in each individual trial. Define $\bm \beta^*$ as the solution of $\mathbb{E}[S_\text{CV}(\bm \beta)]=0$. By the standard asymptotic theory of M-estimation  (cf., \cite{serfling2009approximation}), we can show that $\hat{\bm \beta}_\text{CV} \to \bm \beta^*$, as $n \to \infty$. Since the estimating equations do not account for zero-inflation, there is a discrepancy between $\bm \beta^*$ and the true parameter values $\bm \beta^0$, so the intervention effect estimate from the conventional method, $\hat\beta_{1,\text{CV}}$, is biased. In the current study, we focus on the MLE of the true intervention effect, $\hat\beta_{1,\text{MLE}}$ (defined in Section 2.1.1), which can be recovered by modifying $\hat\beta_{1,\text{CV}}$.

\subsection{Zero-inflation bias}

In this section, we formally describe zero-inflation bias as the difference between the parameters of the true method (i.e., $\beta^0$) and those of the conventional method (i.e., $\beta^*$). Denote $\bm \delta$ as the zero-inflation bias for all parameters, then $\bm \delta=\bm \beta^0-\bm \beta^*$, and $\hat{\bm \beta}_\text{MLE}\approx\hat{\bm \beta}_\text{CV}+\bm \delta$. Since $\hat\beta_{1,\text{MLE}}$ is of primary interest, we focus on the corresponding zero-inflation bias for the intervention effect $\delta_1$ and the following formula $\hat\beta_\text{1,MLE}\approx\hat\beta_\text{1,CV}+\delta_1$.

We can characterize  $\bm \delta$ by taking a close look at the equations $\mathbb{E}[S_{CV}(\bm \beta^*)]=0$. \textcolor{black}{Plugging in $\mathbb{E}[y_i]=(1-\pi_i)\exp({\bf x}_i^t{\bm \beta^0})$,} Equation (\ref{eq:Scv}) can be recast as: 
\begin{equation}
\label{eq:bdelta}
    \begin{split}
        0 = \mathbb{E}[S_{CV}(\bm \beta^*)]=&\,\frac 1n \sum_{i=1}^n\left[(1-\pi_i)\exp\{{\bf x}_i^t(\bm \beta^0-\bm \beta^*)\}-1\right]\exp({\bf x}_i^t\bm \beta^*){\bf x}_i\\
        =&\,\frac 1n \sum_{i=1}^n\{(1-\pi_i)\exp({\bf x}_i^t\bm \delta)-1\}\exp({\bf x}_i^t\bm \beta^*){\bf x}_i\\
        \triangleq&\, \textcolor{black}{B(\bm \delta, \bm \beta^*),}
    \end{split}
\end{equation}
\textcolor{black}{which shows the zero-inflation bias $\bm \delta$ is part of the solution of $B(\bm \delta, \bm \beta^*)=0$.} However, ${\bf x}_i$ and $\pi_i$ require participant-level information, which is unavailable in AD meta-analysis. Hence, Equation (\ref{eq:bdelta}) cannot be solved directly. Alternatively, we can approximate \textcolor{black}{$B(\bm \delta, \bm \beta^*)$} by substituting ${\bf x}_i$ and $\pi_i$ with study-level summary information. 
We describe the approximation in detail in the following section. 

\subsection{Approximate bias $\delta_1$: The ZIBC method}
 
In this section, we describe the ZIBC method to approximate $\delta_1$ using Equation (\ref{eq:bdelta}). 
First, we can simplify \textcolor{black}{$B(\bm \delta, \bm \beta^*)$} by
\begin{equation}
\label{eq:Bapprox}
\begin{split}
\textcolor{black}{B(\bm \delta, \bm \beta^*)} &= \frac 1n \sum_{i=1}^n\{(1-\pi_i)\exp({\bf x}_i^t\bm \delta)-1\}\exp({\bf x}_i^t\bm \beta^*){\bf x}_i\\
&\approx \frac 1n\{(1-\bar \pi)\exp({\bf \bar x}^t\bm \delta)-1\}\sum_{i=1}^n\exp({\bf x}_i^t\bm \beta^*){\bf x}_i\\
&\triangleq \textcolor{black}{C(\bm \delta)D(\bm\beta^*)},
\end{split}
\end{equation}
where \textcolor{black}{$C(\bm \delta)=\frac 1n\{(1-\bar \pi)\exp({\bf \bar x}^t\bm \delta)-1\}$, $D(\bm\beta^*)=\sum_{i=1}^n\exp({\bf x}_i^t\bm \beta^*){\bf x}_i$,} $\bar \pi=\frac{1}{n}\sum_{i=1}^n\pi_i$ is the average structural zero rate, and $\bar{\bf x}$ are the average values for covariates in the sample. Thus, part of the participant-level information (i.e., ${\bf x}_i$ and $\pi_i$) are substituted with the study-level summary statistics (i.e., $\bar{\bf x}$ and $\bar \pi$) to approximate \textcolor{black}{$B(\bm \delta, \bm \beta^*)$}. \textcolor{black}{From the approximation of $B(\bm \delta, \bm \beta^*)\approx C(\bm \delta)D(\bm\beta^*)$, we transfer the problem of solving $B(\bm \delta, \bm \beta^*)=0$ to solving $C(\bm \delta)=0$, which is a function with respect to $\bm \delta$.}

Rewrite ${\bf \bar x} = (1, {\bf \bar z}^t)^t$, where ${\bf \bar z}=( \sum_{i=1}^n{\mathbbm{1}_{\{A_i=T\}}}/n, {\bf \bar x}_{p-2}^t)^t$, and $\bm \delta = (\delta_0, \bm \delta_{p-1}^t)^t$, then \textcolor{black}{$C(\bm \delta)=\frac 1n\{(1-\bar \pi)\exp(\delta_0+ {\bf \bar z}^t\bm \delta_{p-1})-1\}$,
and a solution for $C(\bm \delta)=0$ is}
\begin{equation}
\label{eq:hatdelta}
\hat{\bm \delta}_\text{approx}=\begin{pmatrix} \hat \delta_{0,\text{approx}}  \\  \hat {\bm \delta}_{p-1, \text{approx}}
\end{pmatrix}=\begin{pmatrix} -\log(1-\bar \pi)  \\ {\bf 0}
\end{pmatrix}.
\end{equation}
Thus, the MLE of the true intercept can be recovered by 
\begin{equation}
\label{eq:beta0}
\hat\beta_{0,\text{MLE}} \approx \hat\beta_{0,\text{CV}}+\hat \delta_{0,\text{approx}}= \hat\beta_{0,\text{CV}}-\log(1-\bar \pi).
\end{equation}

\textcolor{black}{Note that the approximation of $B(\bm \delta, \bm \beta^*)$ above is analogous to the 
expectation–maximization (EM) algorithm \citep{dempster1977maximum}, where the expectation-step occurs when plugging in $\mathbb{E}[y_i]$ and average values in Equations (4) and (5). The maximization-step occurs implicitly when maximizing the log-likelihood by taking the first derivative to obtain the estimating equations (i.e., Equation (3)). The advantage of our method is that, after approximation, the estimating equations can be directly solved using summary statistics; therefore, no iterations or IPD are needed. A detailed derivation is provided in Appendix A in the Supplemental Materials.}

However,  $\hat\beta_{1,\text{MLE}}$ cannot be obtained directly as $\hat\delta_{1,\text{approx}}=0$ in Equation (\ref{eq:hatdelta}). To get around this limitation, we can estimate $\hat\beta_{1,\text{MLE}}$ by estimating the MLE of the intercept separately for the control and intervention groups, based on Equation (\ref{eq:beta0}). The specific steps are described as follows:
\begin{enumerate}
\item Consider the sample as being comprised of two separate and independent groups: Intervention and Control.
\item For each group, derive a bias-corrected intercept from the conventional method using Equation (\ref{eq:beta0}). 
\item Merge the corrected intercepts of the two groups from (S2) to obtain the corrected intervention effect estimate. The details are given as follows.
\end{enumerate}
Denote $\mathbb{C}=\{i\,|\,A_i=C, \; i=1,2,...,n\}$ and $\mathbb{T}=\{i\,|\,A_i=T, \; i=1,2,...,n\}$ as the index sets for control and intervention groups, respectively. We further denote $\vert \mathbb{C} \vert = n_C$ and $\vert \mathbb{T} \vert = n_T$. We first consider control group. Since $\mathbbm{1}_{\{A_i=T\}}=0$ for $i\in \mathbb{C}$, Equation (\ref{eq:logmu}) becomes $
\log(\mu_i)=\beta_0+{\bf x}_{i,p-2}^{t}\bm \eta$.
Denote $\hat{\bm \beta}_{C,\text{MLE}}$ and $\hat{\bm \beta}_{C,\text{CV}}$ as the parameter estimates under the true and conventional methods, respectively. Based on Equation (\ref{eq:beta0}), we have 
\begin{equation}
\label{eq:beta0cmle8}
\hat\beta_{0,C, \text{MLE}}
\approx \hat\beta_{0,C,
\text{CV}}-\log(1-\bar{\pi}_C),
\end{equation}
where $\bar \pi_C=\frac{1}{n_C}\sum_{i\in\mathbb{C}}\pi_i$ is the average structural zero rate in control group. For intervention group, since $\mathbbm{1}_{\{A_i=T\}}=1$ for $i\in \mathbb{T}$, Equation (\ref{eq:logmu}) becomes $\log(\mu_i)=\beta_0+\beta_1+{\bf x}_{i,p-2}^{t}\bm \eta$.
Note that the intercept becomes $(\beta_0+\beta_1)$, which includes the intervention effect. Under similar arguments and notations, we then have 
\begin{equation}
\label{eq:beforelemma}
\widehat{(\beta_{0} + \beta_{1})}_{T,\text{MLE}} \approx \widehat{(\beta_{0}+\beta_{1})}_{T,\text{CV}}-\log(1-\bar{\pi}_T),
\end{equation}
where $\hat{\bm \beta}_{T, \text{CV}}$ is the parameter estimate from the conventional method and $\bar \pi_T=\frac{1}{n_T}\sum_{i\in\mathbb{T}}\pi_i$ is the average structural zero rate in intervention group. \textcolor{black}{From Equations (8) and (9), we can see that the discrepancies in intercepts under the true and conventional methods are $-\log(1-\bar{\pi}_C)$ and $-\log(1-\bar{\pi}_T)$ for the control and intervention groups, respectively. Combining the two equations, we can obtain an estimate for the zero-inflation bias $\delta_1$, which is summarized in Lemma 1. The proof is provided in Appendix B in the Supplemental Materials}.

\begin{lemma}
In a study given by Equations (\ref{eq:ZIP}) and (\ref{eq:logmu}), denote the observed covariates excluding the intervention assignment as ${\bf x}_{i,p-2}=(x_{i2},x_{i3},...,x_{i,p-1})^t$ for $i=1,...,n$. If $\bar {\bf x}_{C, p-2}=\bar {\bf x}_{T, p-2}$, where $\bar {\bf x}_{C,p-2}=\frac{1}{n_C}\sum_{i\in \mathbb{C}} {\bf x}_{i, p-2}$ and $\bar {\bf x}_{T,p-2}=\frac{1}{n_T}\sum_{i\in \mathbb{T}} {\bf x}_{i, p-2}$, then we have
\begin{equation}
\label{eq:lemma}
\textcolor{black}{ \hat\beta_{1,\text{MLE}}-\hat\beta_{1,\text{CV}} \approx -\log(1-\bar{\pi}_T)+\log(1-\bar{\pi}_C)\triangleq \hat\delta_1. \;}
\end{equation}
\end{lemma}

\textcolor{black}{We denote the adjusted intervention effect as $\hat\beta_{1,\text{ZIBC}}\triangleq\hat\beta_{1,\text{CV}}+\hat\delta_1$.} Lemma 1 gives the correction formula, Equation (\ref{eq:lemma}), of the proposed ZIBC method. The assumption $\bar {\bf x}_{C, p-2}=\bar {\bf x}_{T, p-2}$ requires that the ``average" subject in control group has the same covariate values as the ``average" subject in intervention group. In a typical two-arm randomized controlled trial, subjects are randomized to either a control or intervention group. Thus the covariates should follow similar distributions across the groups. In addition, the participants in control and intervention groups are expected to be equivalent not only in all measured covariates but also in other unmeasured ones. Hence, the assumption of Lemma 1 can reasonably hold in this case. \textcolor{black}{Note that $\hat\delta_1$ depends on the relative difference between the average structural zero rates of the two groups: $\hat\delta_1 < 0$ if $\bar\pi_T < \bar\pi_C$, $\hat\delta_1 > 0$ if $\bar\pi_T > \bar\pi_C$, and $\hat\delta_1 = 0$ if $\bar\pi_T = \bar\pi_C$. The zero-inflation bias would be minor if the structural zero rates are similar for the control and intervention groups, as the respective influence of zero-inflation cancel each other, even when the zero-inflation itself may be strong. We conducted a simulation study in Section 3 to further evaluate the relationship.} 

The group-level structural zero rates $\bar\pi_C$ and $\bar\pi_T$ can be estimated using the following algorithm. Take the control group, for $i\in\mathbb{C}$, as an example, we have
\begin{equation}
\label{eq:pihat}
\begin{cases}
    \mathbb{E}[\bar y]&=\frac{1}{n_C}\sum_{i \in \mathbb{C}}\mathbb{E}(y_i)=\frac{1} {n_C}\sum_{i \in \mathbb{C}}(1-\pi_i)\mu_i\approx\bar y_{\text{obs}, C},\\
    \mathbb{E}[\text{\# of } y_i=0]&= \sum_{i \in \mathbb{C}}    P(y_i=0)=\sum_{i\in\mathbb{C}}\{\pi_i+(1-\pi_i)e^{-\mu_i}\}\approx n_{0,\text{obs},C},
\end{cases}
\end{equation}
where $n_C$, $\bar y_{\text{obs}, C}$, and $n_{0,\text{obs},C}$ are the sample size, observed outcome average, and observed number of zero outcomes, respectively, for the control group. To estimate $\bar\pi_C$, we approximate Equation (\ref{eq:pihat}) by substituting $\pi_i$ with $\bar\pi_C$, and $\mu_i$ with $\bar \mu_C=\frac{1}{n_C}\sum_{i \in \mathbb{C}}\mu_i$, resulting in 
\begin{equation}
\label{eq:solvepihat}
\begin{cases}
    (1-\bar\pi_C)\bar\mu_C&\approx\bar y_{\text{obs},C},\\
    \{\bar\pi_C+(1-\bar\pi_C)e^{-\bar\mu_C}\}&\approx n_{0,\text{obs},C}/n_C.
\end{cases}
\end{equation}
Here, $n_{0,\text{obs},C}/n_C$ is the proportion of zero outcome values in the control group. By solving Equation (\ref{eq:solvepihat}), we can get an approximation of $\bar\pi_C$. Similarly, we can get $\bar\pi_T$  using the same process. 

The data required for the ZIBC method are (a) $\hat\beta_{1,\text{CV}}$, (b) $\bar y_{\text{obs},C}$, $\bar y_{\text{obs},T}$, and (c) $n_{0,\text{obs},C}/n_C$, $n_{0,\text{obs},T}/n_T$. In a typical trial study, (a) and (b) are directly reported or can be obtained, while (c) are less frequently reported but may be obtained via author queries to the investigators of original studies. 

\subsection{Implementation in meta-analysis}
\label{sec:verify}
Suppose an AD meta-analysis contains $K$ studies that used the conventional method to model zero-inflated outcomes. \textcolor{black}{For each of the $K$ studies, we can apply the ZIBC method to obtain the bias-corrected intervention effect $\hat\beta_{1,\text{ZIBC}}$, which occurs before combining data in a meta-analysis.} For simplicity, we use the reported standard errors $\widehat{\text{SE}}_{1, \text{CV}}$ from the conventional method. With the new set of intervention effects and standard errors, a standard AD meta-analysis can be applied to combine results across studies and obtain the corrected overall intervention effect estimate. \textcolor{black}{For example, a random-effects meta-analysis model, which assumes intervention effects to vary across studies, may be used when study heterogeneity needs to be accounted for in a meta-analysis \citep{dersimonian1986meta, dersimonian2007random}.}

\section{Simulation}
\label{sec:conc}
We conducted simulation studies to examine the performance of the ZIBC method. Specifically, we compare relative performance of the following three methods: 
\begin{enumerate}
\item ZIP regression model (i.e., the true method), the ``gold standard" method, which is not feasible in AD meta-analysis,
\item Poisson regression model (i.e., the conventional method), the method with zero-inflation bias when the outcome is zero-inflated, and
\item ZIBC method, the method to correct zero-inflation bias from the conventional method and recover the intervention effect as if it came from the true method.

\end{enumerate}
In the simulation study, we consider \textcolor{black}{$K=10, 16, \text{and } 20$} randomized clinical trials aimed at evaluating the effect of an intervention on reducing alcohol consumption, where the outcome is the number of standard alcohol drinks. For each trial, we incorporate an additional covariate that follows a standard normal distribution. The simulation was motivated by Project INTEGRATE, a large-scale meta-analysis project examining the effectiveness of brief alcohol interventions on reducing alcohol consumption among young adults \citep{mun2015project}. High proportions of zero alcoholic drinks (i.e., non-drinking) were observed in most trials included in the study. 

The settings of the simulation are based on our observation of the motivating data. Specifically, the sample sizes for individual trials are set at 200 and 400 for first and last half of the studies, respectively. For study $s\in\{1,2,...,K\}$ with sample size $n_s$, the outcome of $i$-th subject ($i\in\{1,2,...,n_s\}$) is simulated by a true ZIP regression model $y_{si}\sim \text{Poisson($\mu_{si}$)}$ with probability $1-\pi_{si}$, and 0 otherwise. The structural zero rate $\pi_{si}$ and Poisson mean parameter $\mu_{si}$ are simulated by $\textrm{logit}(\pi_{si})=\gamma_{0}+\gamma_{1}\mathbbm{1}_{\{A_{si}=T\}}+\gamma_{2}\text{Cov}_{si}$ and $\log(\mu_{si})=\beta_{0}+\beta_{1}\mathbbm{1}_{\{A_{si}=T\}}+\beta_{2}\text{Cov}_{si}$
with a continuous covariate $\text{Cov}_{si}\sim N(0,1)$ and intervention group assignment \textcolor{black}{ $\mathbbm{1}_{\{A_{si}=T\}}\sim\text{Bernoulli}(p^T_s)$, where $p_s^T=0.4, 0.5, 0.6$ for one-third of the studies, respectively, to allow for potential group imbalance.} \textcolor{black}{Note that we will examine $\beta_1$ in the Poisson portion; $\pi_{si}$ is used only to generate data sets, and will not be examined in the simulation study.}

We examine the relative performance of the three methods under the following parameter settings:

1) $(\beta_{0}, \beta_{1}, \beta_{2}) = (1.2, -0.5, 0.25)$,

2) $(\beta_{0}, \beta_{1}, \beta_{2}) = (1.05, -0.35, 0.25)$, and

3) $(\beta_{0}, \beta_{1}, \beta_{2}) = (0.9, -0.2, 0.25)$.\\
Note that as the intervention effect ($\beta_{1}$) varies from $-0.5$ to $-0.2$, the intercept ($\beta_{0}$) also varies accordingly to fix the maximum possible $\log(\mu_{si})$ at the same level of 0.95.

To evaluate the impact of different degrees of zero-inflation on the bias and performance of the methods, we varied the overall proportion of zero drinks at \textcolor{black}{0.2,  0.3, ..., 0.8} among trials. Then $\gamma_{0}, \gamma_{1}$ and $\gamma_{2}$ can be calculated to yield the aforementioned zero rates. In the simulation, we fixed $\gamma_1 = 0.5$, indicating that participants in the intervention group will have a higher probability of no drinking, compared to the control. For example, more participants who previously drank may quit drinking after intervention, compared with their control counterparts. \textcolor{black}{To ensure identifiability of $\gamma_0$, $\gamma_1$, and $\gamma_2$, one additional constraint needs to be applied, and in this simulation, we used $\gamma_{2} = \frac12\gamma_{0}$. Other constraints were considered and examined, and their comparative results from simulation remained the same (results available upon request).}

In one replication of the simulation, data from \textcolor{black}{$K$} intervention studies were generated. For each study, both the true and conventional methods were estimated first, then the ZIBC method was applied to modify the intervention effect estimate from the conventional method. Finally, for each of the three methods, we applied a random-effects meta-analysis model using the \pkg{metafor} R package \citep{viechtbauer2010conducting}, and generated forest plots to compare performance between the methods. 

Figure 1 shows a forest plot from a typical replication during simulation when \textcolor{black}{$K=10$}, true intervention effect $\beta_1=-0.2$, and overall zero rate $=0.4$. Based on the results, we have the following four observations. First, the conventional method produced biased estimates of intervention effects for individual studies as well as the overall result after meta-analysis. Specifically, \textcolor{black}{the estimated zero-inflation bias was positive ($\hat\delta_1=-0.20-(-0.36)=0.16$), as the structural zero rates of intervention groups ($\bar\pi_T$) were higher than those of control groups ($\bar\pi_C$) across the studies, according to Lemma 1.} Second, the true method produced accurate intervention effect estimate, i.e., close to $\beta_1=-0.2$, for each study and the overall effect across studies. Third, the ZIBC method corrected zero-inflation bias to the right direction for each study. Finally, after meta-analysis, the corrected overall estimate from the ZIBC method was very close to the true parameter value of $-0.2$, \textcolor{black}{and the standard error was also close to that of the true method ($0.035$ vs. $0.036$)}. In sum, this typical simulation replication illustrates that the ZIBC method reasonably corrects the biased intervention effect estimates from the conventional method.

\begin{figure}
\begin{center}
\includegraphics[width=5in]{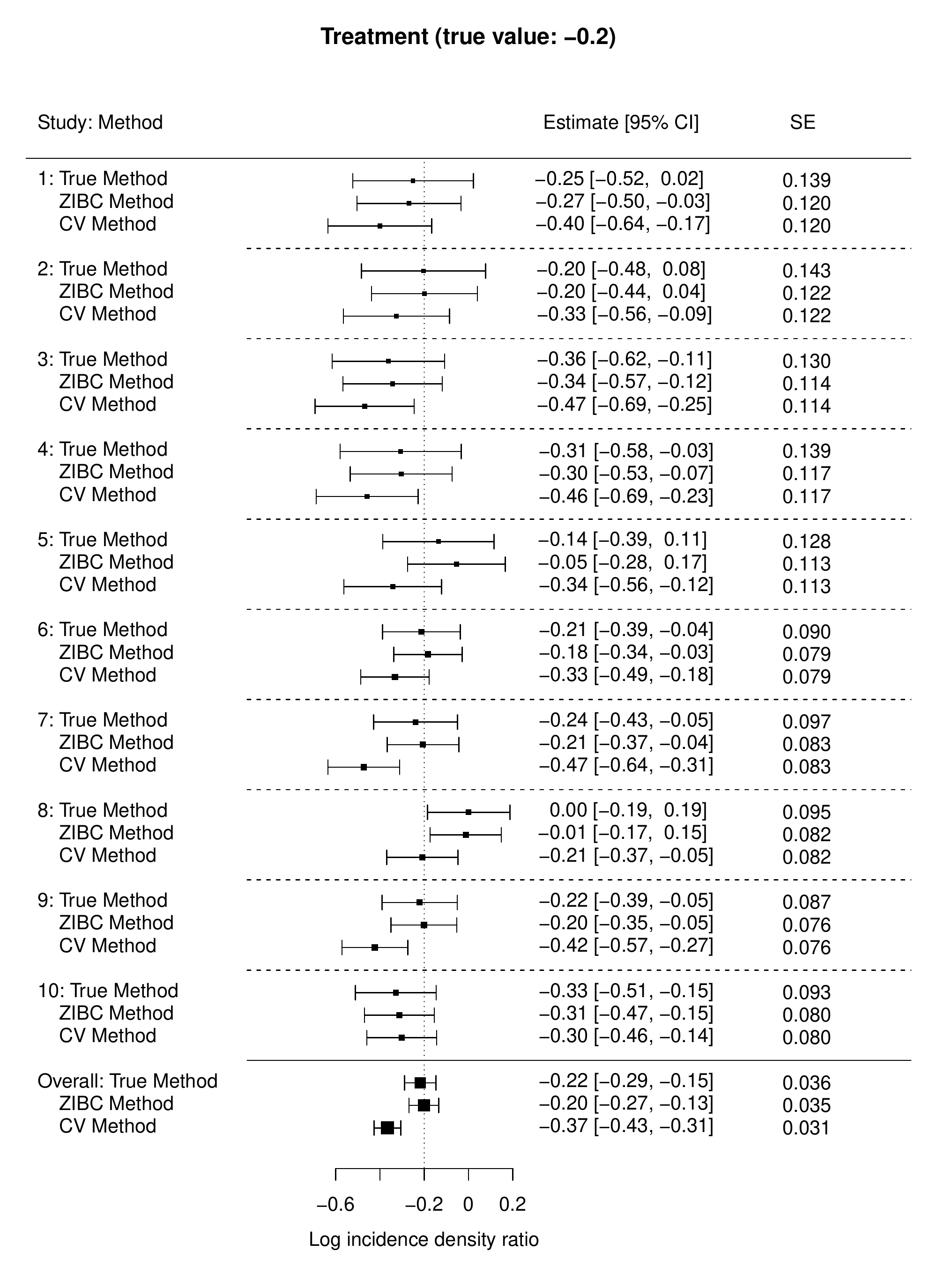}
\end{center}
\caption{A typical forest plot for the true, ZIBC and conventional methods when $\beta_1=-0.2$ \label{fig:1}}
\end{figure}

Figure \ref{fig:1} \textit{graphically} illustrates the good performance of the ZIBC method in a single simulation replication. To examine the performance \textit{numerically} across replications, we compared the intervention effect estimates from the three methods with the true intervention effect $\beta_1$ by calculating the coverage indicator (1 if the 95\% confidence interval covers $\beta_1$ and 0 otherwise) and differences with $\beta_1$ at each replication. After 1000 replications, we calculated the proportion of replications whose 95\% confidence intervals captured $\beta_1$ (coverage rate), and the mean squared error (MSE) between the effect estimate and $\beta_1$. \textcolor{black}{To evaluate the practice of using $\widehat{\text{SE}}_{1, \text{CV}}$ for the ZIBC method in meta-analysis, we calculated the average combined standard errors of the three methods (denoted as average $\widehat{\text{SE}}_{1, \text{MLE}, \text{meta}}$, $\widehat{\text{SE}}_{1, \text{ZIBC}, \text{meta}}$, and $\widehat{\text{SE}}_{1, \text{CV}, \text{meta}}$), as well as the absolute percent relative difference of the conventional method or the ZIBC method against the true method (i.e., $|\widehat{\text{SE}}_{1, \text{CV (or ZIBC)}, \text{meta}} - \widehat{\text{SE}}_{1, \text{MLE}, \text{meta}}|/\widehat{\text{SE}}_{1, \text{MLE}, \text{meta}}$). We compared these indices across the methods.} 

Figure 2 presents the results for different simulation settings \textcolor{black}{when $K=10$. The comparative results when $K=16$ and $20$ (Figure S1 and S2 in Supplemental Materials) are more or less the same as the results of $K=10$.} From the results shown in Figure 2, first, the true method had the highest coverage rates, which were close to 0.95, and also had MSE values close to 0. Second, the conventional method resulted in biased intervention effect estimates, as indicated by low coverage rates and high MSE values. Note that as zero rates increased, zero-inflation bias became greater, leading to progressively lower coverage rates and higher MSE values. Third, the ZIBC method had acceptable coverage rates close to 0.9 and low MSE that were close to 0. Furthermore, the performance of the ZIBC method was consistent across different zero rates \textcolor{black}{between 0.2 and 0.8.} \textcolor{black}{Table 1 presents the average combined standard errors 
and absolute percent relative difference of conventional vs. true method and the ZIBC vs. true method when $K=10$. Compared to the conventional method, the ZIBC method had lower absolute percent relative differences, which were within 3\%, in all scenarios for zero rates $\le 0.7$. For the zero rate of $ 0.8$, absolute percent relative difference of the ZIBC method increased dramatically. This is because as the zero rate approaches to 1, the structural zero rates will also approach to 1, so a small variation in $\bar\pi_C$ (and $\bar\pi_T$) would lead to a more drastic variation in $\log(1-\bar\pi_C)$ (and $\log(1-\bar\pi_T)$) in the correction formula (i.e., Equation (10)). This produces higher standard errors around the parameter estimates. Thus, we recommend using the ZIBC method with caution when the zero rate is 80\% or higher. Based on the comparative results on both the intervention effect estimates and standard errors, the ZIBC method provides reasonable correction for the intervention effect from the conventional method in AD meta-analysis across a wide range of zero inflation.}

\textcolor{black}{We conducted an additional simulation study to further verify the relationship between $\hat\delta_1$ and the relative difference of $\bar\pi_T$ and $\bar\pi_C$ inferred from Lemma 1. Note that the structural zero rates between intervention and control groups are controlled by $\gamma_1$, we, therefore, consider $\gamma_1=-0.5, 0, 0.5$, which represent $\bar\pi_T<\bar\pi_C$, $\bar\pi_T=\bar\pi_C$, and $\bar\pi_T>\bar\pi_C$, respectively, in the simulation. We also consider zero rates of 0.2, 0.4, 0.6, and 0.8, and $(\beta_{0}, \beta_{1}, \beta_{2}) = (1.2, -0.5, 0.25)$. For each pair of $\gamma_1$ and zero rate, under a sample size of 400, Table 2 presents the average $\hat\delta_1$, $\bar\pi_T$, $\bar\pi_C$, $\hat\beta_{1,\text{CV}}$, $\hat\beta_{1,\text{MLE}}$ and $\hat\beta_{1, \text{ZIBC}}$ under 1000 replications. Regardless of the actual zero rates (from 0.2 to 0.8), we observe, on average, $\hat\delta_1<0$ when $\bar\pi_T<\bar\pi_C$ (for $\gamma_1=-0.5$), $\hat\delta_1\approx0$ when $\bar\pi_T\approx\bar\pi_C$ (for $\gamma_1=0$), and $\hat\delta_1>0$ when $\bar\pi_T>\bar\pi_C$ (for $\gamma_1=0.5$), which is consistent with Lemma 1. In addition, we observe that $\hat\beta_{1,\text{CV}}$ is biased when $\gamma_1 \neq 0$, whereas $\hat\beta_{1,\text{ZIBC}}$ is close to $\hat\beta_{1,\text{MLE}}$ in all settings, suggesting that the proposed ZIBC method can provide reasonable correction for the bias  in a wide range of situations.}

\section{Real data analysis}

\subsection{Analysis 1: Project INTEGRATE}
Project INTEGRATE is a large-scale IPD meta-analysis study examining the overall efficacy and comparative effectiveness of brief alcohol interventions for young adults \citep{mun2015project}. A recent IPD meta-analysis of 6,713 participants from 17 randomized controlled trials examined the effect of intervention on the total number of drinks consumed in a typical week, a count variable with a high percentage of zeros  \citep{huh2015brief}. Across all studies, an average of 30\% of individuals reported zero drinking, with the highest proportion of zero drinking being 66\% in one study.  

\textcolor{black}{In this section, we evaluate the performance of the ZIBC method in a real data application. We compared the meta-analysis results between the true, conventional and ZIBC methods using publicly available IPD from Project INTEGRATE \citep{tutorialdata}. As in Section 3 of the simulation
study, IPD were used to estimate parameters from the true and conventional methods. The ZIBC method was conducted using summary statistics from the conventional method (including standard errors for subsequent meta-analysis), mimicking a real data analysis setting where study reports with only summary statistics are available. Intervention studies} included in the current study (a) randomly allocated participants to an intervention or control group, (b) had a follow-up within 6 months from baseline, and (c) had at least one zero outcome in a study. Ten of the 17 studies met the criteria (studies 2, 7 (7.1 and 7.2), 9, 11, 13/14, 15, 16, 18 and 21). For more details of the studies, please refer to \cite{mun2015project}, and \cite{huh2015brief,huh2019tutorial}. The outcome was the average drinks on a typical drinking day in the most recent follow-up assessment within 6 months, with a fixed assessment time for each study. We included the intervention group assignment as the only covariate. 

\textcolor{black}{The comparative results across the three methods are presented in a forest plot (Figure \ref{fig:3}). \textcolor{black}{Since the interventions aimed at reducing the number of alcohol drinks, a negative log incidence rate ratio represents a favorable intervention effect.}}  For most studies, the conventional method produced biased estimates of  intervention effect, compared with the true method. \textcolor{black}{Specifically, the zero-inflation bias was positive (i.e., $\hat\delta_1>0$) in studies 9 and 16, whereas the bias was negative (i.e., $\hat\delta_1<0$) in studies 2, 7.1, 7.2, 15, and 18. In studies 11, 13/14, and 21, the bias  was negligible. Note that although study 11 had very strong zero-inflation ($\bar\pi_T = 67\%$; $ \bar\pi_C=67\%$), because $\bar\pi_T$ was close to $\bar\pi_C$, the zero-inflation bias was very minor. For studies 13/14 and 21, similarly, the zero-inflation bias was small because $\bar\pi_T$ was close to $\bar\pi_C$. This observation is in line with Lemma 1 such that the direction and magnitude of zero-inflation bias depends on the relative difference in structural zero rates between two groups. We also note that the
standard errors from the conventional method were identical up to the second or third decimal place to their counterparts from the true method in each individual study, so the width of the confidence intervals was nearly the same across all three methods. Since the ZIBC method adjusted the effect estimates to the correct level, the confidence intervals of the ZIBC and true methods were nearly the same.} 

The data example demonstrates that the ZIBC method corrects zero-inflation bias regardless of the directions of the bias in the meta-analysis. In conclusion, the ZIBC method showed good performance in correcting zero-inflation bias for \textcolor{black}{study-specific intervention effects as well as the overall pooled intervention effect  in meta-analysis in a real data analysis setting.}

\subsection{Analysis 2: A dental caries prevention clinical trial}

\textcolor{black}{We illustrate the application of the ZIBC method using a randomized controlled trial in dental caries prevention \citep{frazao2011effectiveness}.} The study was aimed at evaluating whether the bucco-lingual technique could increase the effectiveness of a tooth brushing program on preventing dental caries (i.e., cavities) among five-year-old children. This study was a two-arm trial that randomized participants to either a conventional tooth brushing program (Control) or a modified tooth brushing program (Intervention). The outcome of interest was the number of enamel and dentin caries at 18-month follow up, which exhibited considerable zero-inflation, with rates up to 67\%. The conventional Poisson regression model was used to evaluate the intervention effect in the original study. The analysis was stratified by gender due to baseline imbalance in covariates. Since a high proportion of participants did not develop any dental caries, \textcolor{black}{the presence of  zero-inflation bias in the intervention effect estimates from the original study is reasonable to assume. In this example, although IPD were not available, all of the summary information needed to implement the ZIBC method could be extracted from the study report.} 

We apply the ZIBC method here in order to \textcolor{black}{examine the potential} zero-inflation bias. First, we extracted the required information from the original study (see Table 3). Specifically, the uncorrected effects  (i.e., $\hat \beta_{1, \text{CV}}$ and  $\widehat{\text{SE}}_{1, \text{CV}}$) were calculated from incidence density ratios (IDR) and 95\% confidence intervals in the original Table 3 from \cite{frazao2011effectiveness}, and the arm-level outcome averages and the proportion of zeros (i.e., $\bar y_{\text{obs},C}$, $\bar y_{\text{obs},T}$, $n_{0,\text{obs},C}/n_C$ and $n_{0,\text{obs},T}/n_T$) were obtained directly from the original Figure 2 by using software WebPlotDigitizer version 4.2 \citep{rohatgi2018webplotdigitizer}. We then estimated the arm-level average structural zero rates $\bar \pi_C$ and $\bar \pi_T$ by solving Equation (\ref{eq:solvepihat})
, which were 49\% and 32\%, respectively, for girls, and 27\% and 45\%, respectively, for boys. Finally, we obtained the corrected intervention effect estimates $\hat \beta_{1, \text{MLE}}$ by plugging the values of $\hat \beta_{1, \text{CV}}$, $\bar \pi_C$ and $\bar \pi_T$ into Equation (\ref{eq:lemma}). Using the original standard errors $\widehat{\text{SE}}_{1, \text{CV}}$, we obtained the modified p-values based on the Wald test. 

The original and ZIBC method-corrected results are summarized in Table 4.  \textcolor{black}{According to the original analysis, girls receiving the modified tooth brushing program tended to develop more caries with an IDR of $1.34$, suggesting a potentially negative or harmful intervention effect.}  After applying the ZIBC method to adjust for the zero-inflation, the IDR was corrected to $1.01$, \textcolor{black}{suggesting a null intervention effect. Note that the intervention effect was statistically insignificant before and after the correction, so the statistical conclusion did not change after applying the ZIBC method. For boys, the original analysis reported a significant protective intervention effect with an IDR of $-0.74$ (P-value = 0.02)}. After applying the ZIBC method, the intervention effect was reduced to an IDR of $-0.46$ and \textcolor{black}{became statistically insignificant (P-value = 0.13), suggesting that the original statistical conclusion may not be valid.}

In meta-analysis, the statistical significance of an intervention effect in an individual study is less important than its magnitude and uncertainty, which can influence the overall pooled result. Therefore, adjusting biased effect sizes would improve precision in drawing statistical inference. When evaluating the effect of the tooth brushing program on dental caries, it would be better to utilize bias corrected estimates rather than estimates from the original report. Namely, 0.01 for girls and $-$0.46 for boys. Standard errors can also be taken from the original study (i.e., 0.28 for girls; 0.30 for boys) because we found standard errors from the original study can reasonably substitute unknown standard errors associated with the bias-corrected intervention effect estimates (see Table 1 and Figure 3).
\section{Discussion and conclusion}
\textcolor{black}{In this paper, we propose the ZIBC method to correct zero-inflation bias that may arise in the intervention effect estimates of clinical trials with excessive zero outcome values in AD meta-analysis. Specifically, this method aims to recover the intervention effect estimates from a conventional Poisson model as if they were appropriately estimated in a ZIP model.} 
The ZIBC method works well when one can use the information of the ``average" subject in the sample to approximate the study result, as we substitute IPD required in the estimating equations with their group-level average values to relax the IPD requirement. \textcolor{black}{The idea of substituting IPD with average values is in line with the Mean Value Theorem for Integrals and the EM algorithm}. 
The statistical property of the ZIBC method is justified by Lemma 1, which is based on the assumption that the characteristics (or covariates) of ``average" subjects in \textcolor{black}{control and intervention} groups are similar, which should hold in randomized controlled trials due to random assignment to groups. In other situations where the assumption is not met, such as case-control or cross-sectional studies, the ZIBC method should be used with caution. In addition, by imposing linear predictors in the true ZIP regression model (i.e., Equation (\ref{eq:logmu})), we implicitly assume no intervention by covariate interactions on the outcome, which should hold in most trials. \textcolor{black}{We note that the \textit{intervention effect} targeted by the ZIBC method is the mean difference between two groups and cannot be interpreted as a \textit{causal effect} \citep{zheng2020causal}. If one is interested in drawing causal inference, then issues of noncompliance \citep{frangakis2002principal} and assumptions of temporal stability, causal transience, and unit homogeneity \citep{holland1986statistics} need to be taken into consideration.}

\textcolor{black}{The adjusted intervention effect estimates from the ZIBC method correspond to the Poisson portion in the ZIP model, characterizing the subpopulation that may or may not engage in the targeted behavior, which may be of greater interest in certain meta-analyses. In contrast, the intervention effect estimates derived from the conventional Poisson model pertain to the entire population.  However, the ZIBC method can be used to incorporate these studies into a meta-analysis focusing on the subpopulation described above.  
In practice, we recommend that researchers check the population of interest before applying the ZIBC method when conducting meta-analysis. 
We also acknowledge the ideal way for combining studies with presumably biased effects is to communicate with the original investigators and request IPD, so that meta-analysts can re-analyze raw data using statistical methods that are most suited to the research question. However, IPD may not be available due to data sharing restriction  and other resource limitations.  
The proposed ZIBC method can serve as a practical alternative to adjust for zero-inflation bias in an AD meta-analysis when obtaining original data is not feasible.}

\textcolor{black}{In data analysis, having a high proportion of zeros does not necessarily mean that zero-inflation bias exists in the estimated intervention or treatment effect size. The ZIBC method should be considered only when a proportion of zeros in data exceeds the expected proportion given a Poisson parameter. For example, when the mean of a Poisson distribution is equal to 1, the expected zero rate is 36.8\%. This high rate of zeros would be in line with the Poisson model when the average value of the outcome is low in quantity (e.g., 1) and there would be no need for the ZIBC method even if the actual zero rate were as high as 40\%.  We recommend that the ZIBC method be used when the actual zero rate is much higher than the one expected when fitting a Poisson model. In another example, the mean number of drinks following an alcohol intervention would be usually much higher than 1 (e.g., 3). For Poisson distribution with a mean of 3, its corresponding expected zero rate is less than 5.0\%. Therefore, an actual zero rate of 20\% or higher would signal a need to account for zero-inflation bias. Additionally, in the specific context of bias correction for an intervention effect size estimate, as illustrated by Lemma 1, zero-inflation bias may not occur when the intervention and control groups have similar zero rates, even when zero rates in both groups are high (e.g., study 11 in Project INTEGRATE). In situations where there is a difference in zero rates between groups, we recommend that the ZIBC method be used. Note that the Poisson model is nested within the ZIP model, so misspecifying the ZIP model when the Poisson model is accurate will not lead to biased estimates but will result in efficiency loss due to the estimation of additional parameters. The consequence of incorrectly  specifying a ZIP model when data follow a Poisson distribution is relatively minor, while the opposite would lead to a biased estimate. Therefore, when the proportion of observed zeros is considerably higher than what was expected or when there is a difference in the proportions of zeros between groups, the ZIBC method can be considered.}


\textcolor{black}{The ZIBC method adjusts the intervention effect for each of the studies separately and independently, which occurs before combining data for meta-analysis.} After correcting any zero-inflation bias for each individual trial, modified intervention effects are then combined in AD meta-analysis to obtain a more accurate overall result.  \textcolor{black}{Note that the ZIBC method only targets the mean intervention effect estimates, corresponding to a first-order correction. It would be theoretically attractive to adjust  standard  errors for  zero  inflation  bias as well, a second-order correction. However, it is beyond the scope of the current study and can be investigated in future studies. For simplicity, we used the standard errors from the Poisson models when conducting AD meta-analysis, which showed
reasonable performance in our simulation study and real data examples.}

\textcolor{black}{The ZIBC method minimally requires summary information for its correction. In many situations, all the required data can be directly obtained from study reports (e.g., the real data example in Section 4.2).} It also requires the group-level outcome zero rates, which \textcolor{black}{sometimes may not be described in study reports} but can be obtained through inquiries with original investigators, or an educated guess when prior information or expert knowledge is available. \textcolor{black}{
Note that the outcome average and zero rate are sufficient statistics for a ZIP distribution, so they are good substitutes for IPD  when only AD are available.} 

The ZIBC method we describe can be extended in the future in several ways. First, although we illustrate the ZIBC method in the context of a two-arm trial design, it can be applied to multi-arm trials by comparing each intervention group with control and correcting the biased intervention effect per pair. Second, aside from the ZIBC method, alternative strategies may be investigated for their feasibility and validity when adjusting the estimating equations for zero-inflation bias. One potential strategy is to generate pseudo IPD based on AD of outcome and each covariate, and then solve for $\delta$ using the pseudo data, which is similar to the idea of Approximate Bayesian Computing (see, e.g., \cite{marin2012approximate, beaumont2002approximate}).  Finally, the proposed method is designed to recover biased intervention effect estimates from the conventional Poisson model when the ZIP regression model should have been used; however, it can be extended to other statistical models with appropriate adjustments, such as a negative binomial regression model and a two-sample \textit{t}-test, which can be thought of as a Wald test in a simple linear regression with intervention group membership as the lone covariate. 


\section*{Funding}
This work was supported by National Institutes of Health grants (R01
AA019511, K02 AA028630) and National Science Foundation grants (DMS1737857, 1812048, 2015373 and 2027855). The content is solely the responsibility
of the authors and does not necessarily represent the official views of the NIAAA, the
National Institutes of Health, or the NSF.

\section*{Data Availability Statement}
The data from Project INTEGRATE used in this paper to illustrate our findings are openly available in Mendeley at http://doi.org/10.17632/4dw4kn97fz.2 \citep{tutorialdata}.

\bibliography{ZIBC_sim}

@article{mun2015project,
  author = {Mun, Eun-Young and {de la Torre}, Jimmy and Atkins, David C. and White, Helene R. and Ray, Anne E. and Kim, Su-Young and Jiao, Yang and Clarke, Nickeisha and Huo, Yan and Larimer, Mary E. and Huh, David},
  title = {Project {INTEGRATE}: An integrative study of brief alcohol interventions for college students.},
  journal = {Psychology of Addictive Behaviors},
  volume = {29},
  number = {1},
  pages = {34-48},
  ISSN = {1939-1501 (Electronic)
0893-164X (Linking)},
  DOI = {10.1037/adb0000047},
  howpublished = "\url{http://www.ncbi.nlm.nih.gov/pubmed/25546144}",
  year = {2015},
  type = {Journal Article}
}

@article{sutton2008recent,
  title={Recent developments in meta-analysis.},
  author={Sutton, Alexander J and Higgins, Julian PT},
  journal={Statistics in Medicine},
  volume={27},
  number={5},
  pages={625--650},
  year={2008},
  publisher={Wiley Online Library}
}

@article{lyman2005strengths,
  title={The strengths and limitations of meta-analyses based on aggregate data.},
  author={Lyman, Gary H and Kuderer, Nicole M},
  journal={BMC Medical Research Methodology},
  volume={5},
  number={1},
  pages={14},
  year={2005},
  publisher={BioMed Central}
}

@article{huh2015brief,
  title={Brief motivational interventions for college student drinking may not be as powerful as we think: An individual participant-level data meta-analysis.},
  author={Huh, David and Mun, Eun-Young and Larimer, Mary E and White, Helene R and Ray, Anne E and Rhew, Isaac C and Kim, Su-Young and Jiao, Yang and Atkins, David C},
  journal={Alcoholism: Clinical and Experimental Research},
  volume={39},
  number={5},
  pages={919--931},
  year={2015},
  publisher={Wiley Online Library}
}

@article{huh2019tutorial,
  title={A tutorial on individual participant data meta-analysis using Bayesian multilevel modeling to estimate alcohol intervention effects across heterogeneous studies.},
  author={Huh, David and Mun, Eun-Young and Walters, Scott T and Zhou, Zhengyang and Atkins, David C},
  journal={Addictive Behaviors},
  volume={94},
  pages={162--170},
  year={2019},
  publisher={Elsevier}
}

@article{marin2012approximate,
  title={Approximate Bayesian computational methods.},
  author={Marin, Jean-Michel and Pudlo, Pierre and Robert, Christian P and Ryder, Robin J},
  journal={Statistics and Computing},
  volume={22},
  number={6},
  pages={1167--1180},
  year={2012},
  publisher={Springer}
}

@article{beaumont2002approximate,
  title={Approximate Bayesian computation in population genetics.},
  author={Beaumont, Mark A and Zhang, Wenyang and Balding, David J},
  journal={Genetics},
  volume={162},
  number={4},
  pages={2025--2035},
  year={2002},
  publisher={Genetics Soc America}
}

@article{viechtbauer2010conducting,
  title={Conducting meta-analyses in R with the metafor package.},
  author={Viechtbauer, Wolfgang},
  journal={Journal of Statistical Software},
  volume={36},
  number={3},
  pages={1--48},
  year={2010},
  publisher={UCLA Statistics}
}

@article{garcia2004trial,
  title={A trial of antiparasitic treatment to reduce the rate of seizures due to cerebral cysticercosis.},
  author={Garcia, H{\'e}ctor H and Pretell, E Javier and Gilman, Robert H and Martinez, S Manuel and Moulton, Lawrence H and Del Brutto, Oscar H and Herrera, Genaro and Evans, Carlton AW and Gonzalez, Armando E},
  journal={New England Journal of Medicine},
  volume={350},
  number={3},
  pages={249--258},
  year={2004},
  publisher={Mass Medical Soc}
}

@article{silcocks2010p3mc,
  title={P3MC: A double blind parallel group randomised placebo controlled trial of Propranolol and Pizotifen in preventing migraine in children.},
  author={Silcocks, Paul and Whitham, Diane and Whitehouse, William Patrick},
  journal={Trials},
  volume={11},
  number={1},
  pages={71},
  year={2010},
  publisher={BioMed Central}
}

@article{frazao2011effectiveness,
  title={Effectiveness of the bucco-lingual technique within a school-based supervised toothbrushing program on preventing caries: a randomized controlled trial.},
  author={Fraz{\~a}o, Paulo},
  journal={BMC Oral Health},
  volume={11},
  number={1},
  pages={11},
  year={2011},
  publisher={Springer}
}

@article{rohatgi2018webplotdigitizer,
  title={WebPlotDigitizer Version: 4.2.},
  author={Ankit Rohatgi},
  journal={https://automeris.io/WebPlotDigitizer},
  year={2019}
}

@book{serfling2009approximation,
  title={Approximation theorems of mathematical statistics},
  author={Serfling, Robert J},
  year={1980},
  publisher={John Wiley \& Sons}
}

@article{murphy2000comparative,
  title={Comparative hospitalization of hemodialysis and peritoneal dialysis patients in Canada.},
  author={Murphy, Sean W and Foley, Robert N and Barrett, Brendan J and Kent, Gloria M and Morgan, Janet and Barr{\'e}, Paul and Campbell, Patricia and Fine, Adrian and Goldstein, Marc B and Handa, S Paul and others},
  journal={Kidney International},
  volume={57},
  number={6},
  pages={2557--2563},
  year={2000},
  publisher={Elsevier}
}

@article{long2014marginalized,
  title={A marginalized zero-inflated Poisson regression model with overall exposure effects.},
  author={Long, D Leann and Preisser, John S and Herring, Amy H and Golin, Carol E},
  journal={Statistics in Medicine},
  volume={33},
  number={29},
  pages={5151--5165},
  year={2014},
  publisher={Wiley Online Library}
}

@article{lambert1992zero,
  title={Zero-inflated Poisson regression, with an application to defects in manufacturing.},
  author={Lambert, Diane},
  journal={Technometrics},
  volume={34},
  number={1},
  pages={1--14},
  year={1992},
  publisher={Taylor \& Francis}
}

@book{schmid2020handbook,
  title={Handbook of Meta-Analysis},
  author={Schmid, Christopher H and Stijnen, Theo and White, Ian},
  year={2020},
  publisher={CRC Press}
}

@article{chen2020relative,
  title={Relative efficiency of using summary versus individual data in random-effects meta-analysis.},
  author={Chen, Ding-Geng and Liu, Dungang and Min, Xiaoyi and Zhang, Heping},
  journal={Biometrics},
  year={2020},
  publisher={Wiley Online Library}
}

@incollection{liu2018avenues,
  title={Avenues for further research},
  author={Liu, Yulun and Chen, Yong},
  booktitle={Diagnostic Meta-Analysis },
  pages={305--315, },
  year={2018},
  publisher={ Springer}
}

@article{dempster1977maximum,
  title={Maximum likelihood from incomplete data via the EM algorithm.},
  author={Dempster, Arthur P and Laird, Nan M and Rubin, Donald B},
  journal={Journal of the Royal Statistical Society: Series B (Methodological)},
  volume={39},
  number={1},
  pages={1--22},
  year={1977},
  publisher={Wiley Online Library}
}

@article{zheng2020causal,
  title={Causal inference in randomized clinical trials.},
  author={Zheng, Cheng and Dai, Ran and Gale, Robert Peter and Zhang, Mei-Jie},
  journal={Bone Marrow Transplant},
  pages={4--8},
  year={2020},
  publisher={Springer}
}

@article{frangakis2002principal,
  title={Principal stratification in causal inference.},
  author={Frangakis, Constantine E and Rubin, Donald B},
  journal={Biometrics},
  volume={58},
  number={1},
  pages={21--29},
  year={2002},
  publisher={Wiley Online Library}
}

@article{holland1986statistics,
  title={Statistics and causal inference.},
  author={Holland, Paul W},
  journal={Journal of the American Statistical Association},
  volume={81},
  number={396},
  pages={945--960},
  year={1986},
  publisher={Taylor \& Francis}
}

@article{dersimonian1986meta,
  title={Meta-analysis in clinical trials.},
  author={DerSimonian, Rebecca and Laird, Nan},
  journal={Controlled Clinical Trials},
  volume={7},
  number={3},
  pages={177--188},
  year={1986},
  publisher={Elsevier}
}

@article{dersimonian2007random,
  title={Random-effects model for meta-analysis of clinical trials: an update.},
  author={DerSimonian, Rebecca and Kacker, Raghu},
  journal={Contemporary Clinical Trials},
  volume={28},
  number={2},
  pages={105--114},
  year={2007},
  publisher={Elsevier}
}

@article{zeileis2008regression,
  title={Regression models for count data in R.},
  author={Zeileis, Achim and Kleiber, Christian and Jackman, Simon},
  journal={Journal of Statistical Software},
  volume={27},
  number={8},
  pages={1--25},
  year={2008},
  publisher={Foundation for Open Access Statistics}
}

@article{kelly2017pilot,
  title={A pilot randomized clinical trial testing integrated 12-Step facilitation (iTSF) treatment for adolescent substance use disorder.},
  author={Kelly, John F and Kaminer, Yifrah and Kahler, Christopher W and Hoeppner, Bettina and Yeterian, Julie and Cristello, Julie V and Timko, Christine},
  journal={Addiction},
  volume={112},
  number={12},
  pages={2155--2166},
  year={2017},
  publisher={Wiley Online Library}
}

@article{milgrom2009xylitol,
  title={Xylitol pediatric topical oral syrup to prevent dental caries: a double-blind randomized clinical trial of efficacy.},
  author={Milgrom, Peter and Ly, Kiet A and Tut, Ohnmar K and Mancl, Lloyd and Roberts, Marilyn C and Briand, Kennar and Gancio, Mary Jane},
  journal={Archives of Pediatrics \& Adolescent Medicine},
  volume={163},
  number={7},
  pages={601--607},
  year={2009},
  publisher={American Medical Association}
}

@article{horton2007cautionary,
  title={A cautionary note regarding count models of alcohol consumption in randomized controlled trials.},
  author={Horton, Nicholas J and Kim, Eugenia and Saitz, Richard},
  journal={BMC Medical Research Methodology},
  volume={7},
  number={1},
  pages={1--9},
  year={2007},
  publisher={BioMed Central}
}

@misc{tutorialdata,
  author = {Huh, David and Mun, Eun-Young and Walters, Scott T and Zhou, Zhengyang and Atkins, David C},
  title = {Data and code for: A tutorial on individual participant data meta-analysis using Bayesian multilevel modeling to estimate alcohol intervention effects across heterogeneous studies.},
  year = {2019},
  howpublished = {Mendeley Data, V2}, 
  url = {http://dx.doi.org/10.17632/4dw4kn97fz.2}
}

\bibliographystyle{agsm}
\newpage

\begin{table}[H]
\centering
 \caption{Comparison of standard errors among the true, ZIBC and conventional methods ($K=10$)}
 	\vspace{0.3cm}
\begin{tabular}{|ccccccc|}
\hline
\multirow{2}{*}{$\beta_1$}&\multirow{2}{*}{Zero rate} & Average & Average & Average & APRD & APRD \\
 &                        & $\widehat{\text{SE}}_{1, \text{MLE}, \text{meta}}$ &  $\widehat{\text{SE}}_{1, \text{ZIBC}, \text{meta}}$ & $\widehat{\text{SE}}_{1, \text{CV}, \text{meta}}$  &  ZIBC vs. True &  CV vs. True \\ \hline
             & 0.2        & 0.029 & 0.029 & 0.028 &    0.003     &  0.047         \\
  & 0.3        & 0.032 & 0.031 & 0.032 &     0.021    &  0.005         \\
  & 0.4        & 0.035 & 0.034 &  0.038&      0.016    & 0.090          \\
 $-0.2$ & 0.5        & 0.039 & 0.038 & 0.046 &    0.011     & 0.196          \\
  & 0.6        & 0.044 & 0.044 & 0.057 &    0.007     &  0.283         \\
  & 0.7        & 0.053 & 0.053 & 0.071 &    0.001     & 0.341          \\
  & 0.8        & 0.068 & 0.093 & 0.091 &   0.363      &   0.332        \\ \hline
 & 0.2        & 0.029 & 0.030 & 0.028  &   0.007      &  0.052         \\
 & 0.3        & 0.033 & 0.032 & 0.032 &    0.025     &   0.024        \\
 & 0.4        & 0.036 & 0.035 & 0.038 &    0.019     &    0.061       \\
$-0.35$ & 0.5        & 0.040  & 0.039 & 0.046 &   0.014     & 0.164           \\
 & 0.6        & 0.045 & 0.045 & 0.057 &    0.012     &   0.254        \\
 & 0.7        & 0.054 & 0.056 & 0.071 &    0.031     &   0.312        \\
 & 0.8        & 0.070  &  0.081 & 0.090 &  0.163         &  0.300         \\ \hline
  & 0.2        & 0.030  & 0.031 & 0.028 &    0.022     &  0.052         \\
  & 0.3        & 0.034 & 0.033 & 0.032 &   0.029      &  0.045         \\
  & 0.4        & 0.037 & 0.036 & 0.038 &   0.024      &  0.033         \\
$-0.5$  & 0.5        & 0.041 & 0.040 &  0.046 &   0.020       &  0.130          \\
  & 0.6        & 0.046 & 0.046 &  0.057 &   0.015       &   0.220        \\
  & 0.7        & 0.056 & 0.057  & 0.070 &    0.022     &  0.267         \\
  & 0.8        & 0.071 &  0.082 & 0.090 &    0.154     & 0.268 \\   \hline
\end{tabular}
{\raggedright APRD = Absolute percent relative difference. \par}
\end{table}
\newpage

\begin{table}[H]
\centering
 \caption{Empirical verification of Lemma 1 with 1000 simulations ($\beta_1=-0.5$)}
 	\vspace{0.3cm}
\begin{tabular}{|cccccccc|}
\hline
\multirow{2}{*}{$\gamma_1$}&\multirow{2}{*}{Zero rate} & Average&Average&Average&Average&Average&Average\\
 &                        & $\hat\delta_1$ &  $\bar\pi_T$ & $\bar\pi_C$ &  $\hat\beta_{1,\text{CV}}$ &  $\hat\beta_{1,\text{MLE}}$ &  $\hat\beta_{1, \text{ZIBC}}$ \\ \hline
           & $0.2$        & $-0.045$ & $0.119$ & $0.155$ & $-0.460$ &  $-0.505$  &  $-0.502$     \\
\multirow{2}{*}{$-0.5$ }     & $0.4$        & $-0.162$ & $0.300$ & $0.400$  & $-0.346$ &  $-0.508$ &    $-0.499$     \\
     & $0.6$        & $-0.277$ & $0.505$ & $0.617$ & $-0.230$ &  $-0.507$  &  $-0.488$     \\
     & $0.8$        & $-0.359$ & $0.736$ & $0.810$ & $-0.154$ &  $-0.514$  &  $-0.488$     \\\hline
       & $0.2$        & $-0.001$ & $0.144$ & $0.135$ & $-0.504$ &  $-0.505$ &   $-0.494$     \\
\multirow{2}{*}{$0$ } & $0.4$   & $0.000$ & $0.358$ & $0.350$ & $-0.506$ &  $-0.507$ &   $-0.493$     \\
       & $0.6$        & $ 0.003$ & $0.568$ & $0.562$ &  $-0.512$ &  $-0.509$ & $-0.496$     \\
       & $0.8$        & $0.003$ & $0.778$ & $0.775$ & $-0.523$ &  $-0.520$  &  $-0.511$     \\\hline
      & $0.2$        & $0.039$ & $0.166$ & $0.115$ & $-0.545$ &  $-0.505$  &  $-0.484$      \\
\multirow{2}{*}{$0.5$}      & $0.4$        & $0.153$ & $0.414$ & $0.302$ & $-0.660$ &  $-0.507$ &   $-0.483$     \\
      & $0.6$        & $0.282$ & $0.631$ & $0.506$ & $-0.795$ &  $-0.513$  &  $-0.501$     \\
      & $0.8$        & $0.391$ & $0.821$ & $0.738$ & $-0.910$  & $-0.519$ &   $-0.522$\\  \hline
\end{tabular}
\end{table}
\newpage

\begin{table}[H]
\centering
 \caption{Information extracted from \cite{frazao2011effectiveness}}
 	\vspace{0.3cm}
\begin{tabular}{|cccc|}
\hline
Summary information & Data source                        & \;Girls\; &\; Boys\; \\ \hline
      $\hat \beta_{1, \text{CV}}$              & Table 3                        & 0.29          & $-0.73$        \\ 
      $\widehat{\text{SE}}_{1, \text{CV}}$               & Table 3                        & 0.28          & 0.30         \\ 
      $\bar y_{\text{obs},C}$              & Figure 2 (with WebPlotDigitizer) & 0.83          & 1.04         \\ 
     $\bar y_{\text{obs},T}$               & Figure 2 (with WebPlotDigitizer) & 1.06          & 0.49         \\ 
     $n_{0,\text{obs},C}/n_C$               & Figure 2 (with WebPlotDigitizer) & 59\%          & 45\%         \\ 
      $n_{0,\text{obs},T}/n_T$              & Figure 2 (with WebPlotDigitizer) & 47\%          & 67\%         \\ \hline
\end{tabular}
\end{table}
\newpage
\begin{table}[H]
\centering
 \caption{Original and ZIBC method-corrected intervention effect estimates, incidence density ratios (IDRs) and P-values, for girls and boys, respectively}
 \vspace{0.3cm}
\begin{tabular}{|ccccc|}
\hline
                       &           & \;Estimate\; & \;\; IDR \;\;  & \;P-value\; \\ \hline
\multirow{2}{*}{\;\; Girls\;\; } & Original  & 0.29     & 1.34 & 0.29    \\
                       & Corrected & 0.01     & 1.01 & 0.97    \\ \hline
\multirow{2}{*}{\; Boys \;}  & Original  & $-0.73$    & 0.48 & 0.02    \\
                       & Corrected & $-0.46$    & 0.63 & 0.13    \\ \hline
\end{tabular}
\end{table}

\setlength{\tabcolsep}{.16667em}

\begin{figure}
\begin{center}
\includegraphics[width=6in]{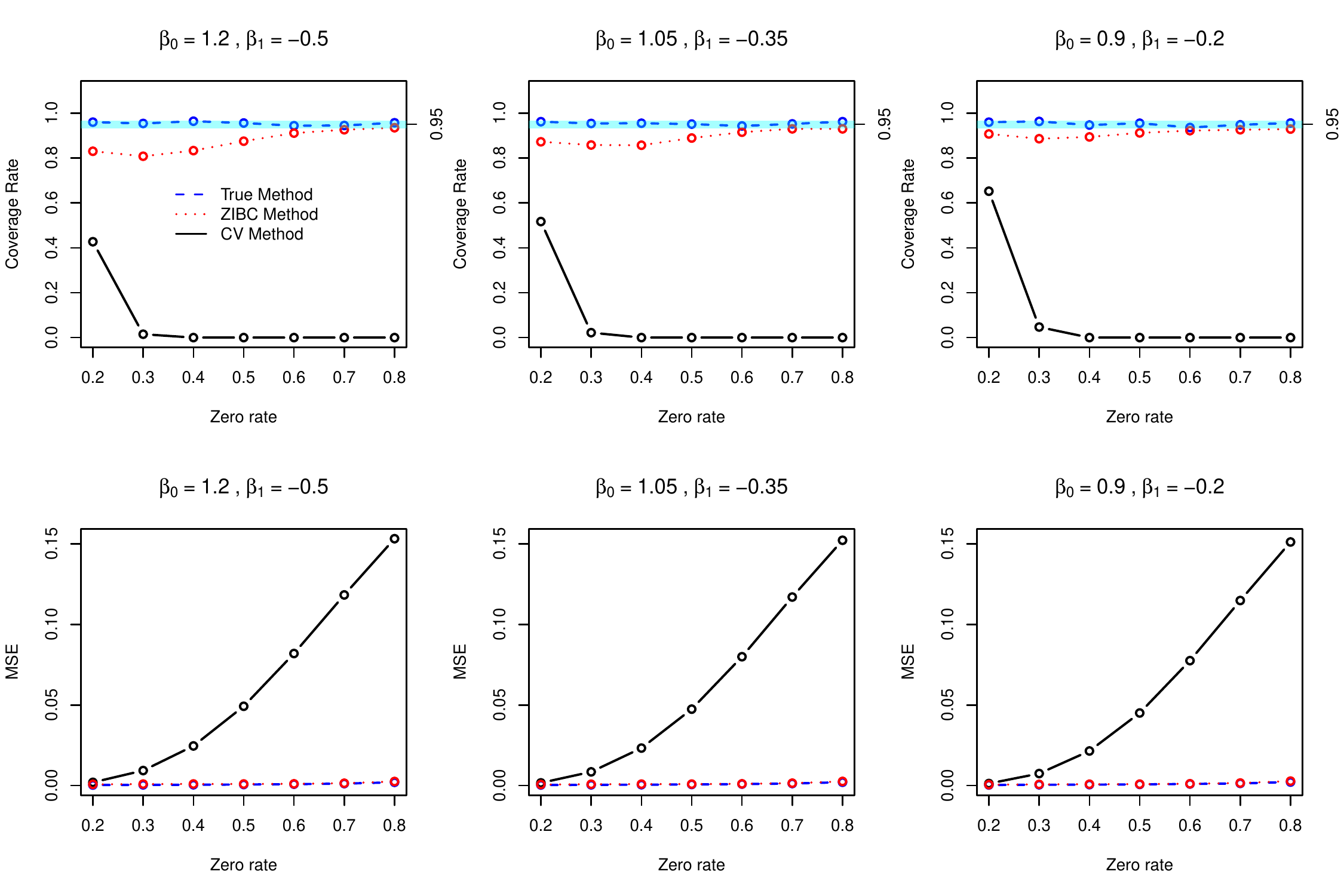}
\end{center}
\caption{Coverage rates and MSE values of the true (blue dashed line), ZIBC (red dotted line) and conventional (black solid line) methods from 1000 replications ($K=10$)\label{fig:2} }
\end{figure}


\begin{figure}
\begin{center}
\includegraphics[width=4in]{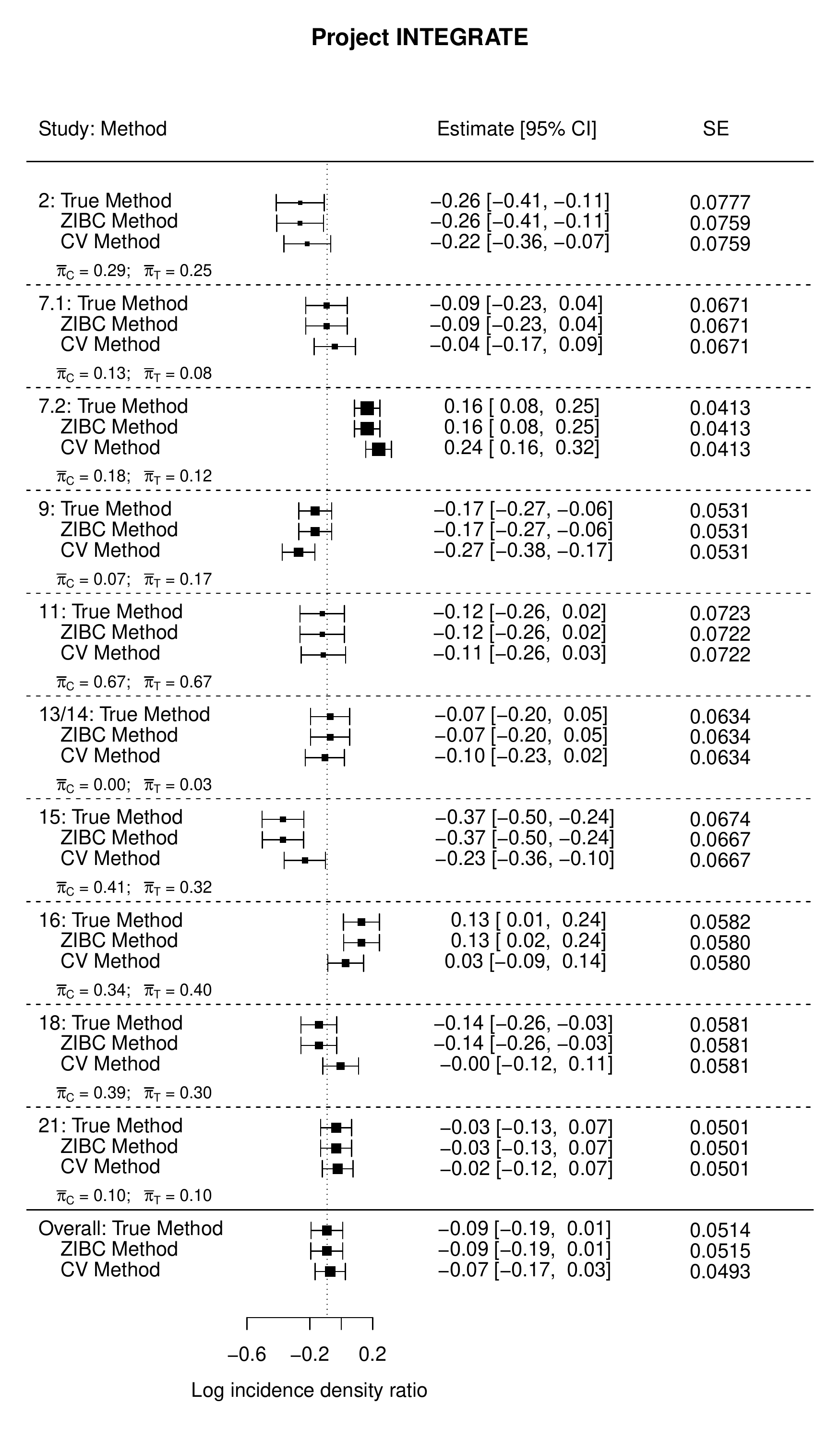}
\end{center}
\caption{Forest plot for the true, ZIBC and conventional methods in Project INTEGRATE ($\hat\beta_{1, \text{MLE}}=-0.09$, $\hat\beta_{1, \text{ZIBC}}=-0.09$, $\hat\beta_{1, \text{CV}}=-0.07$ after meta-analysis) \label{fig:3}}
\end{figure}

\newpage

\end{document}


\begin{center}
		{\large \bf A Bias Correction Method in Meta-analysis of Randomized Clinical Trials with no Adjustments for Zero-inflated Outcomes: Supplementary Materials
		}
		
		\vspace{.7 cm}
		
		Zhengyang Zhou, Ph.D. \\
		Department of Biostatistics and Epidemiology \\
		University of North Texas Health Science Center, Fort Worth, TX \\
		
		\vspace{.5 cm}
		Minge Xie, Ph.D. \\
		Department of Statistics\\
		Rutgers University,
		Piscataway, NJ \\

		\vspace{.5 cm}
		David Huh, Ph.D.  \\
		School of Social Work \\
		University of Washington,	Seattle, WA\\
		
	    \vspace{.5 cm}
		Eun-Young Mun, Ph.D.  \\
		Department of Health Behavior and Health Systems \\
		University of North Texas Health Science Center, Fort Worth, TX \\
		
	\end{center}
	
	\vspace{1cm}
	
	\noindent
	Correspondence should be sent to:\\
		Zhengyang Zhou, Ph.D. \\
		Email: zhengyang.zhou@unthsc.edu 
		   \vspace{.5 cm}
		   \\
		Eun-Young Mun, Ph.D. \\
		Email: eun-young.mun@unthsc.edu \\

\section*{Appendix A. Analogue between the ZIBC method and EM algorithm}
 We can obtain  Equation (5) following the idea of the EM algorithm. Denote $\log L_{CV}({\bm \beta})\triangleq \sum_{i=1}^n\{-\exp({\bf x}_i^t{\bm \beta})+y_i{\bf x}_i^t{\bm \beta} \}$ as the log-likelihood function of ${\bm \beta}$ under conventional model minus a constant, the analogous expectation and maximization steps are shown as follows. 
\begin{itemize}
\item (Analogous) expectation-step:\\
Taking the expectation of $\log L_{CV}({\bm \beta})$, we have
\begin{equation*}
\begin{split}
\mathbb{E}[\log L_{CV}({\bm \beta})] &= \sum_{i=1}^n\{-\exp({\bf x}_i^t{\bm \beta})+\mathbb{E}[y_i]\cdot{\bf x}_i^t{\bm \beta} \}\\
&=\sum_{i=1}^n\{-\exp({\bf x}_i^t{\bm \beta})+(1-\pi_i)\exp({\bf x}_i^t{\bm \beta^0})\cdot{\bf x}_i^t{\bm \beta} \}\\
&\approx \sum_{i=1}^n\{-\exp({\bf x}_i^t{\bm \beta})+(1-\bar\pi)\exp(\bar{\bf x}^t{\bm \beta^0})\cdot{\bf x}_i^t{\bm \beta} \}\\
&\triangleq Q({\bm \beta}).
\end{split}
\end{equation*}
\end{itemize}

\begin{itemize}
\item (Analogous) maximization step: \\
To maximize $Q({\bm \beta})$, we derive its first derivative with respect to ${\bm \beta}$, which is given by
\begin{equation*}
\begin{split}
\frac{\partial Q({\bm \beta})}{\partial{\bm \beta}}&=\sum_{i=1}^n\{-\exp({\bf x}_i^t{\bm \beta})+(1-\bar\pi)\exp(\bar{\bf x}^t{\bm \beta^0})\}{\bf x}_i\\
&=\left[(1-\bar\pi)\exp(\bar{\bf x}^t\{{\bm \beta^0}-{\bm \beta}\})-1\right]\sum_{i=1}^n\exp({\bf x}_i^t{\bm \beta}){\bf x}_i.
\end{split}
\end{equation*}
\end{itemize}
Note that ${\bm \beta^*}$ is the solution of $S_\text{CV}(\bm \beta)=0$. By plugging ${\bm \beta^*}$ in $\frac{\partial Q({\bm \beta})}{\partial{\bm \beta}}$, we obtain Equation (5).

\newpage

\section*{Appendix B. Proof of Lemma 1}
\setcounter{equation}{0}
\renewcommand{\theequation}{A.\arabic{equation}}
\begin{proof}
Consider three ``average" subjects in the control group, intervention group, and the overall sample (denoted as $\{\text{average},C\}$, $\{\text{average},T\}$, and $\{\text{average}\}$), with ${\bf x}_{\text{average},C,p-2} = \bar {\bf x}_{C, p-2}$, ${\bf x}_{\text{average},T,p-2} = \bar {\bf x}_{T, p-2}$, and ${\bf x}_{\text{average},p-2} = \bar {\bf x}_{p-2}$, respectively. Without loss of generality, assuming that observed covariates excluding the intervention assignment are grand mean centered before data analysis, we have ${\bf x}_{\text{average},p-2}=\bar {\bf x}_{p-2}=\bf 0$. Since $\bar {\bf x}_{C, p-2}=\bar {\bf x}_{T, p-2}$, we also have 
${\bf x}_{\text{average},C,p-2}={\bf x}_{\text{average},T,p-2}={\bf x}_{\text{average},p-2}=\bf 0$. Therefore, we have
\begin{equation}
\nonumber
\begin{split}
&\log(\mu_{\text{average},C})=\beta^0_{0,C}\\
&\log(\mu_{\text{average},T})=\beta^0_{0,T}+\beta^0_{1,T}\\
&\log(\mu_{\text{average}})=\beta^0_{0}+\beta^0_1\mathbbm{1}_{\{A_\text{average}=T\}}
\end{split}
\end{equation}
under the true method. If an average subject in the overall sample belongs to the control group, then
\begin{equation}
\label{eq:beta0c}
\begin{split}
&\log(\mu_{\text{average}})=\log(\mu_{\text{average},C})\\
\Rightarrow& \beta^0_{0,C}=\beta^0_{0}\\
\Rightarrow& \hat\beta_{0, C, \text{MLE}} \approx\hat\beta_{0, MLE}.
\end{split}
\end{equation}
Similarly, if an average subject belongs to the intervention group, then
\begin{equation}
\label{eq:betatmle}
\begin{split}
&\log(\mu_{\text{average}})=\log(\mu_{\text{average},T})\\
\Rightarrow& \beta^0_{0,T}+\beta^0_{1,T}=\beta^0_{0}+\beta^0_{1}\\
\Rightarrow& \widehat{(\beta_{0}+\beta_{1})}_{T,\text{MLE}}\approx\hat\beta_{0, MLE}+\hat\beta_{1, MLE}.
\end{split}
\end{equation}
Under similar arguments, for the conventional method, we have 
\begin{equation}
\label{eq:beta0ccv}
\hat\beta_{0,C,\text{CV}}\approx \hat\beta_{0,\text{CV}}
\end{equation}
and
\begin{equation}
\label{eq:betatcv}
\widehat{(\beta_{0}+\beta_{1})}_{T,\text{CV}}\approx\hat\beta_{0, CV}+\hat\beta_{1, CV}.
\end{equation}
Plug Equations (\ref{eq:beta0c}) and (\ref{eq:beta0ccv}) into Equation (8), and plug Equations (\ref{eq:betatmle}) and (\ref{eq:betatcv}) into Equation (9), we have
\begin{equation}
\begin{split}
&\hat\beta_{0, \text{MLE}}
\approx \hat\beta_{0,
\text{CV}}-\log(1-\bar{\pi}_C)\\
&\hat\beta_{0,\text{MLE}} + \hat\beta_{1,\text{MLE}} \approx \hat\beta_{0,\text{CV}}+\hat\beta_{1,\text{CV}}-\log(1-\bar{\pi}_T),
\end{split}
\end{equation}
which directly gives Equation (10).
\end{proof}

\newpage
 \renewcommand{\thefigure}{S1}
\begin{figure}
\begin{center}
\includegraphics[width=6in]{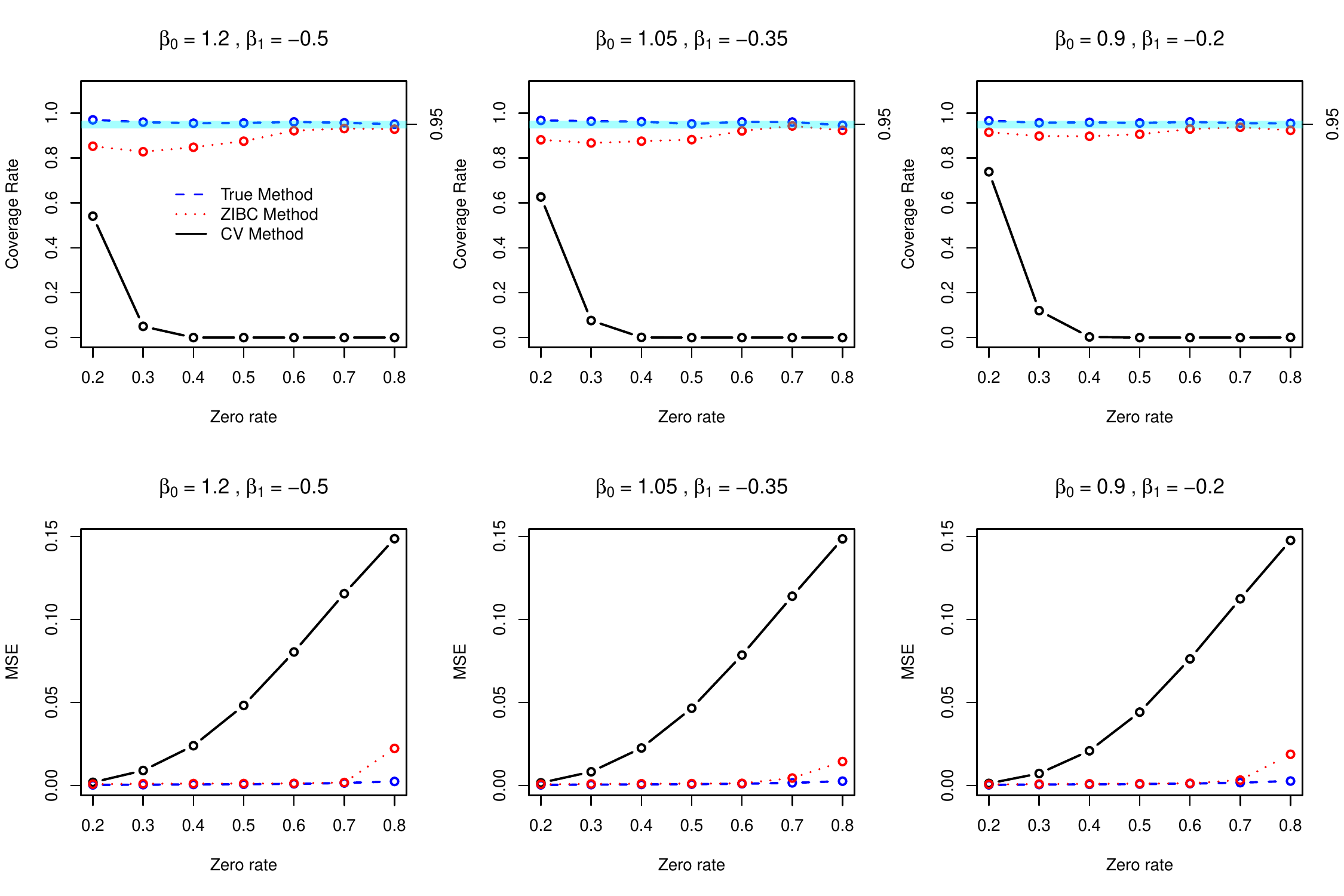}
\end{center}
\caption{Coverage rates and MSE values of the true (blue dashed line), ZIBC (red dotted line) and conventional (black solid line) methods from 1000 replications (K=16)\label{fig:S1} }
\end{figure}

 \renewcommand{\thefigure}{S2}
\begin{figure}
\begin{center}
\includegraphics[width=6in]{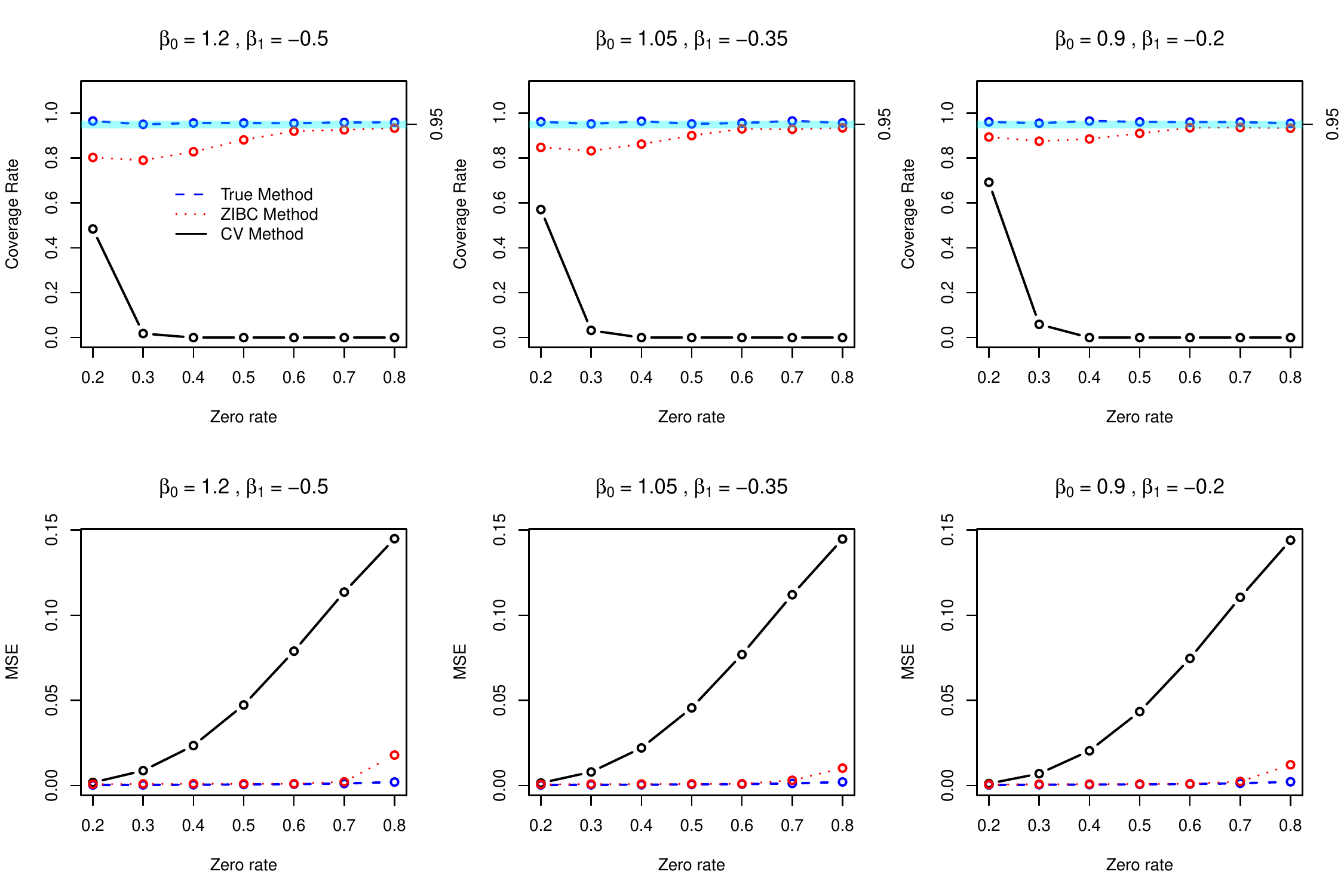}
\end{center}
\caption{Coverage rates and MSE values of the true (blue dashed line), ZIBC (red dotted line) and conventional (black solid line) methods from 1000 replications (K=20)\label{fig:S2} }
\end{figure}